\documentstyle[graphicx,12pt,aasms4]{article}

\def\lsim{\raise0.3ex\hbox{$<$}\kern-0.75em{\lower0.65ex\hbox{$\sim$}}}
\def\gsim{\raise0.3ex\hbox{$>$}\kern-0.75em{\lower0.65ex\hbox{$\sim$}}}

\def\kms{\rm ~km~s^{-1}}

\begin{document}

\title{WIND-INTERACTION MODELS FOR THE EARLY AFTERGLOWS OF GAMMA-RAY BURSTS: 
\break THE CASE OF GRB 021004}

\author{Zhi-Yun Li and Roger A. Chevalier}
\affil{Department of Astronomy, University of Virginia, P.O. Box 3818}
\affil{Charlottesville, VA 22903; zl4h@virginia.edu, rac5x@virginia.edu}


\begin{abstract}

Wind-interaction models for gamma-ray burst afterglows predict that 
the optical emission from the reverse shock drops below that from 
the forward shock within 100s of seconds of the burst. The typical 
frequency $\nu_m$ of the synchrotron emission from the forward shock 
passes through the optical band typically on a timescale of minutes 
to hours. Before the passage of $\nu_m$, the optical flux evolves 
as $t^{-1/4}$ and after the passage, the decay steepens to 
$t^{-(3p-2)/4}$, where $p$ is the exponent for the assumed power-law 
energy distribution of nonthermal electrons and is typically $\sim 2$. 
The steepening in the slope of temporal decay should be readily 
identifiable in the early afterglow light curves. We propose that 
such a steepening was observed in the R-band light curve of GRB 021004 
around day 0.1. Available data at several radio frequencies are 
consistent with this interpretation, as are the X-ray observations 
around day~1. The early evolution of GRB 021004 contrasts with that 
of GRB 990123, which can be described by emission from interaction 
with a constant density medium.

\end{abstract}


\section{INTRODUCTION}

The initial model for the afterglows of gamma-ray bursts (GRBs) 
involved synchrotron emission from nonthermal electrons accelerated 
to a power-law spectrum in a relativistic spherical blast wave 
expanding into a constant-density, presumably interstellar, medium 
(ISM; M\'esz\'aros \& Rees 1997). This simplest model 
has difficulty, however, explaining quantitatively the dozen or so 
sources whose
afterglows are observed well enough to allow for detailed modeling
(e.g., Panaitescu \& Kumar 2002). 
The most commonly discussed complication
 is a collimated energy injection (Rhoads 1997; Sari, Piran 
\& Halpern 1999). This jet model provides a reasonable fit to the 
majority of the well observed afterglows, assuming a constant-density 
ambient medium (Panaitescu \& Kumar 2002). For some sources, a 
blast wave expanding into an ambient medium of $r^{-2}$ 
density distribution, as expected of a stellar wind, can fit the 
data equally well or even better (Chevalier \& Li 2000, CL00 hereafter; 
Li \& Chevalier 2001; Panaitescu \& Kumar 2002). 
Notable examples include GRB 970508 (see 
Frail, Waxman \& Kulkarni 2000 for a different view) and GRB 011121 
(Price et al. 2002). There is evidence for two types of GRB ambient  
environments, with implications for their progenitors. They are not 
immediately distinguishable because, at an age of a few days, the 
preshock wind density is comparable to an interstellar density. At 
earlier times, the density contrast is higher and the jet effects 
are less important. The early afterglow observations are expected 
to provide a better handle on the nature of the ambient medium. 

In this Letter, we summarize the characteristics of the early 
afterglows expected in the wind model (\S~\ref{windmodel}), and 
argue that the afterglow observations of GRB 021004 are consistent 
with the source interacting with a Wolf-Rayet type wind 
(\S~\ref{grb021004}). The strongest 
evidence for wind interaction comes from the initial slow decay 
of the R-band light curve and its prominent steepening around 
day 0.1. 

\section{ANALYTIC LIGHT CURVES OF EARLY AFTERGLOWS}
\label{windmodel}

Analytic light curves for the standard ISM model are given in Sari, Piran 
\& Narayan (1998), assuming a power-law electron energy spectrum with 
index $p$ and constant fractions of blast wave energy, $\epsilon_e$ 
and  $\epsilon_B$, going into nonthermal electrons and magnetic field
respectively. At any given time, the synchrotron spectrum is determined
by a set of characteristic frequencies: the typical frequency $\nu_m$,
cooling frequency $\nu_c$, and self-absorption frequency $\nu_a$. 
The light curve at any given frequency $\nu$ is determined by the 
characteristic times $t_m$, $t_c$ and $t_a$, when $\nu_m$, $\nu_c$ 
and $\nu_a$ cross $\nu$, and the critical time $t_0$ when $\nu_m$ 
and $\nu_c$ become equal. The light curve of the ISM model was 
extended to the wind case by CL00 (see also Panaitescu \& Kumar 2000 
and Granot \& Sari 2002). In the wind model, the cooling and self-absorption 
frequencies are expected to be lower at early times than 
those in the ISM model, because of a higher ambient density at small 
radii. The cooling frequency has a strong effect on the early 
emission in the optical and the self-absorption frequency in radio. 

At the earliest times, the optical flux may be dominated by the emission from
the reverse shock front, but the forward shock front is not much
fainter for wind interaction (CL00, eq. [58]), assuming that the two
shocks have similar efficiencies for the production of synchrotron radiation.
Once the reverse shock front has passed through the initial shell,
which is expected to occur on a timescale similar to that of the
gamma-ray burst, the reverse shock
emission drops sharply because it is in the fast cooling regime.
Unless the reverse shock is ``refreshed'' by a continued flow, the
decline is determined by off-axis emission that arrives at a later
time (Kumar \& Panaitescu 2000).
Optical frequencies are expected to be between $\nu_c$ and $\nu_m$,
so the flux is $\sim \nu^{-1/2}$ and decays as $t^{-5/2}$
(Kumar \& Panaitescu 2000).

To describe the forward shock emission, we use the characteristic
times mentioned above, which were estimated by CL00 assuming $p=2.5$. 
We rescale the 
estimates to the R-band with $\nu_R=(\nu/4.5\times 10^{14}$~Hz), and 
obtain 
\begin{equation}
t_m=0.04 (1+z)^{1/3}\epsilon_{e,-1}^{4/3}\epsilon_{B,-1}^{1/3}
E_{52}^{1/3} \nu_R^{-2/3}\ \ {\rm days},
\label{tm}
\end{equation}
which is about an hour for standard parameters. The parameter $z$ is
the cosmological redshift, $E_{52}$ is the blast wave energy in
units of $10^{52}$~ergs, and $\epsilon_n=\epsilon/10^n$. 
Note that $t_m$ does not depend on the wind 
density $A$ directly, where $\rho=Ar^{-2}$, 
although the wind must be sufficiently dense that the 
transition from fast cooling to slow cooling occurs after $t_m$. It 
depends most sensitively on $\epsilon_e$. The condition on $\epsilon_e$ 
for the R-band break to occur between one minute and one hour is
$ 
0.005  
< \epsilon_e (1+z)^{1/4}\epsilon_{B,-1}^{1/4}E_{52}^{1/4}< 0.1,
$ 
which covers a reasonable range, and has only a weak dependence on other 
parameters. The transition to slow cooling occurs around the time 
$
t_0=1(1+z)\epsilon_{e,-1}\epsilon_{B,-1} A_*\  {\rm day},
$
which is typically later than $t_m$ in the R-band. 
Here, $A_*=A/5\times 10^{11}{\rm g~cm^{-1}}$. 
The cooling frequency 
crosses the frequency $\nu_R$ at a time 
$
t_c=5\times 10^3 (1+z)^3 \epsilon_{B,-1}^3 E_{52}^{-1} A_*^4 \nu_R^2\ \ {\rm days},
$
which is typically later than both $t_m$ and $t_0$. The self-absorption
frequency $\nu_a$ is typically well below the optical, and can be
ignored. 

Before $t_0$, the synchrotron electrons are in the fast cooling regime, 
and the flux peaks at the cooling frequency $\nu_c$, so $F_{\nu_c}=F_{\nu,max}$
with
\begin{equation} 
F_{\nu,max}=2.1 
	(1+z)^{3/2} \epsilon_{B,-1}^{1/2} E_{52}^{1/2}A_* d_{L1}^{-2} 
	t_{\rm days}^{-1/2}\ \ 
	{\rm mJy}, 
\label{Fnumax}
\end{equation}
where $ d_{L1}$ is the luminosity distance in units of 10 Gpc.
The flux at the typical frequency $\nu_m$ is lower, and is 
given by
\begin{equation} 
F_{\nu_m}=2.7
	(1+z)^{1/2}\epsilon_{e,-1}^{-1} \epsilon_{B,-1}^{-1/2} 
	E_{52}^{1/2} d_{L1}^{-2} 
	t_{\rm days}^{1/2}\ \ {\rm mJy},
\label{Fnum_1}
\end{equation}
which  is independent of the wind density.
After the transition to slow cooling at $t_0$, the flux peaks at $\nu_m$ 
rather than $\nu_c$, and now $ F_{\nu_m}=F_{\nu,max}$ (eq. [\ref{Fnumax}]).
There is a general
scaling for the late-time R-band flux after the break ($t > t_m$)  
$
F_{\nu_R}(t)\propto \epsilon_e^{p-1} \epsilon_B^{(p-2)/4}E_{52}
^{(p+2)/4} t^{-(3p-2)/4} \propto t_{m,R}^{3(p-1)/4} \epsilon_B
^{-1/4} E_{52}^{3/4} t^{-(3p-2)/4},  
$   
where $t_{m,R}$ is the break time at the R-band. 
Sources with later breaks tend to be brighter.

\section{MODELING GRB 021004}
\label{grb021004}

The gamma-ray burst GRB 021004 was detected by the HETE II satellite 
 (Shirasaki et al. 2002) and had
an optical afterglow detected within minutes of the $\gamma$-ray 
burst (Fox et al. 2003b). 
The early light curves of the only other afterglows detected at such
early times,
GRBs 990123 (Akerlof et al. 1999)
and 021211 (Fox et al. 2003a), have similar shapes, both showing a 
rapid initial decline followed by a slower decay, although GRB 
021211 is fainter by about 3 magnitudes at similar epochs. 
In both sources, the
initial rapidly decaying emission is interpreted as coming from 
the reverse shock of GRB ejecta running into a constant density
medium (Sari \& Piran 1999; Li et al. 2003; Fox et al. 2003a; 
Wei 2003). The 
early optical afterglow of GRB 021004 shows a different behavior: 
it has a very slow initial decay of $t^{-0.4\pm 0.1}$, followed 
by a steepening around day~0.1 into approximately $t^{-1}$ 
(Fox et al. 2003b). Kobayashi \& Zhang (2003) interpreted the 
early afterglow data in terms of emission from a combination of 
reverse and forward shocks expanding into a constant-density 
medium. Fox et al. (2003b) questioned this interpretation, and
suggested instead a continued energy injection into the blast wave 
after the $\gamma$-ray burst to explain the initial slow 
decay. We propose that the slow decay is a natural consequence 
of the fast cooling ($\nu_c < \nu_m$) expected in a wind model 
at early times and that the steepening is caused by the typical 
frequency $\nu_m$ passing through the optical band from above 
while still in the fast cooling regime. We show  that this 
interpretation, besides fitting the R-band light curve, is in a 
reasonable agreement with the radio data available at several 
frequencies.

The free parameters that 
appear in the wind model can be estimated analytically
using the formulae given in the 
previous section. Our identification of the break time $t_{m,R}=
0.1$~day in the R-band light curve yields, using eq. 
(\ref{tm}), the relation 
$ 
\epsilon_{e,-1}^{4/3}\epsilon_{B,-1}^{1/3}
E_{52}^{1/3} =1.7,
$ 
for a redshift of $z=2.32$ (Chornock \& Filippenko 2002). 
For a cosmological model with $H_0=71\kms~{\rm Mpc^{-1}}$,
$\Omega_M=0.27$, and $\Omega_{\Lambda}=0.73$, this redshift
corresponds to $d_{L1}=1.89$.
At the 
break, the R-band flux is approximately $F_{\nu_m}=0.83$~mJy, 
which yields a second relation
$ 
\epsilon_{e,-1}^{-1} \epsilon_{B,-1}^{-1/2} E_{52}^{1/2}=1.9
$ 
from eq.~(\ref{Fnum_1}). The wind density $A_*$ does
not enter into either of the two relations, which enables us to
express $\epsilon_e$ and $E_{52}$ in terms of $\epsilon_B$:
$ 
\epsilon_e =0.11\epsilon_{B,-1}^{-1/3}\ \ 
{\rm and} \ \  E_{52}=4.0 \epsilon_{B,-1}^{1/3}.
$  

To constrain the wind density $A_*$, we note that the typical frequency 
$\nu_m$ decreases with time as $t^{-3/2}$. It should cross the 8.46 GHz 
wavelength around day 140. This is much later than the time $t_0$ for 
transition to slow cooling, which occurs around
$  
t_0=3.7 \epsilon_{B,-1}^{2/3} A_*\ \ {\rm days},
$  
for typical parameters. Therefore, we can use eq.~(\ref{Fnumax}) 
to find the expected peak flux at 8.46 GHz. The result is 
\begin{equation}
F_{\nu_m,8.46 {\rm GHz}}=0.59 \epsilon_{B,-1}^{2/3} 
	A_* \ \ {\rm mJy},
\label{c5}
\end{equation}
which for standard parameters is comparable to the R-band flux at the 
break $t_{m,R}$. 
The 8.46 GHz flux is observed at 598$\pm$33~$\mu$Jy on day 5.7. If this
flux is comparable to the peak flux at the time $t_m$ (which is true 
if the observed frequency is close to or beyond the self-absorption
frequency; see CL00), then one can use eq.~(\ref{c5}) to provide a 
rough estimate for $A_*$ in terms of $\epsilon_B$:
$ 
A_*\approx 1.0 \epsilon_{B,-1}^{-2/3}.
$ 

The above analytic estimates assumed $p=2.5$, and are rather crude.
They do indicate that the optical and radio data may be fitted
with a wind model with parameters not far from the standard values. 
We now demonstrate that this is indeed the case using a numerical 
model. The model treats synchrotron emission from a spherical 
(trans-)relativistic blast wave propagating in an $r^{-2}$ density 
medium, with the light travel time effects, synchrotron self 
absorption and cooling included. It was previously applied to GRB 
980508 (CL00), among others. Like GRB 980508, the decay of the R-band 
flux on the time scale of days and longer is relatively 
slow, with $F_\nu\propto t^{-1}$ approximately; the wiggles on 
the light curve of this source makes a precise determination of the 
decay slope difficult. The slope implies that $p$ is close to 2, 
although the exact value is somewhat uncertain. We pick $p=2.1$,  
which corresponds to a decay slope of $\alpha=-(3p-2)/4=1.075$ in  
the optical. 

After some experimentation, we find a solution that fits the R-band 
and radio data reasonably well with the following set of parameters: 
$\epsilon_e=0.1$, $\epsilon_B=0.1$, $E_{52}=10$, and $A_*=0.6$,
which corresponds to a
wind mass loss rate of $6\times 10^{-6}$~M$_\odot$~yr$^{-1}$ (assuming 
a nominal wind speed of $10^3$~km~s$^{-1}$).
Now all the parameters are approximately determined (within a factor
$\sim 2$) because of the inclusion of self-absorption.
The fits are shown in panels (a) and (b) of Fig.~1. We did not 
attempt to fit the bumps on the R-band light curve; they have been 
interpreted as arising from either late energy injections or 
inhomogeneities in the ambient density (Lazzati et al. 
2002; Nakar, Piran \& Granot 2003; Heyl \& Perna 2003). The bumps
introduce some uncertainty to the model parameters we obtained. Radio 
emission was detected at 4.86, 8.46, 15, 22.5 and 86 GHz at various times
(Frail \& Berger 2002; Berger, Frail \& Kulkarni 2002; Pooley 2002b,c;
Bermer \& Castro-Tirado 2002). One upper limit each exists at 15, 232 
and 347 GHz (Pooley 2002a; Bermer \& Castro-Tirado 2002; Wouterloot 
et al. 2002). The flux measurement 
of 2.5$\pm$0.3 mJy at 86 GHz flux at an average time of 1.5~days
is particularly interesting. This flux is three times higher than 
the R-band flux at the break around day 0.1. It presents a problem to 
the identification of the break as $t_m$ in a {\it constant density} 
medium when the cooling frequency has already passed the R-band 
from below (Kobayashi \& Zhang 2002). In such a case, the maximum 
fluxes at lower frequencies should be the same as that of the R-band
at the break, namely about 0.83~mJy, which is well below the 86 GHz
measurement. This discrepancy was also noted by Pandey et al. (2002). 

The relatively high 86 GHz flux is not a problem for our  model,
where the transition to slow cooling occurs around day 2, much later 
than day~0.1. 
In a wind model, the flux can be much higher in the radio (broadly
defined to include millimeter and sub-millimeter wavelengths), 
particularly at early times when the cooling frequency is expected 
to be in the spectral region. This behavior shows up clearly in 
panel (b) of Fig.~1, where the peak fluxes at the three highest
frequencies are all above 3 mJy. Such high fluxes are naturally 
expected in a wind model but not in an ISM model, as emphasized
by Panaitescu \& Kumar (2000). However, it is difficult to make a
strong case for wind interaction based on a single data point at 
86 GHz. A stronger case can be made if the 8.46 GHz flux starts to 
decline around 100~days, when the typical frequency $\nu_m$ is 
expected to pass through the frequency from above. This expectation
needs to be modified in the case of an early jet break.

X-ray afterglows are observed with Chandra at two epochs. The first
epoch started about 0.87 days after the burst and lasted for
88.1 ksec (Sako \& Harrison 2002a; Fox et al. 2003b). Within this
epoch, the X-ray afterglow has a power-law spectrum, with index 
$\beta_X=-1.1\pm 0.1$, and decays roughly as a power-law, with 
index $\alpha_X=-1.0\pm 0.2$. Both are consistent with our model, 
where the cooling frequency $\nu_c$ around day 1 is well below the
X-ray band, and $\beta_X$ and $\alpha_X$ are predicted to be --1.05
and --1.075, respectively, for $p=2.1$. The fact that the temporal 
decay slope in the R-band, $\alpha_O$, is close to --1 indicates 
that the cooling frequency is below the R-band as well around this 
time, which is in agreement with the optical spectral index of 
$\beta_O=-1.07\pm 0.06$ determined by Pandey et al. (2002) and 
$\beta_O=-0.96\pm 0.03$ by Matheson et al. (2003) in the absence 
of a substantial host galaxy extinction (see also Bersier et al. 
2003; Holland et al. 2002). Our best fit model yields an X-ray 
flux of $2.6\times 10^{-13}$~erg~cm$^{-2}$~s$^{-1}$ between 2 and 
10 keV at day 1.4, close to 
the middle of the first observing epoch. It is lower than, but 
within a factor of two of, the mean flux of the entire epoch $4.3
\times 10^{-13}$~erg~cm$^{-2}$~s$^{-1}$ (Sako \& Harrison 2002a). 
We therefore conclude that the wind model is consistent with the 
first epoch of X-ray observations. 

The second epoch of X-ray observations started 52.23 days 
after the burst and yielded a 
2-10 keV flux of 7.2$\pm$2.5$\times 10^{-16}$ erg~cm$^{-2}$~s$^
{-1}$ (Sako \& 
Harrison 2002b). The flux implies a decay slope between the two epochs of 
approximately $\alpha_X=-1.7$, which is steeper than that predicted 
in our spherical model. The steepening may be due to a jet break
between the two epochs of observation, which was suggested to have
occurred around day 7 by Pandey et al. (2002) and Holland et al.
(2002) based on their interpretation of the (wiggly) R-band 
light curve. The jet break, if exists, should show up in a well-sampled
radio light curve as well. Alternatively, the steepening could 
be due to a steepening of the energy distribution of nonthermal
electrons well above the minimum energy of the electrons accelerated
at the shock front (e.g., Li \& Chevalier 2001; Panaitescu \& Kumar 
2002). 


The question of a jet break is related to the energy in the source.
The energy we find in a spherical model for the afterglow, $E_{52}=10$,
is comparable to the isotropic burst energy in $\gamma$-rays,
$5\times 10^{52}$ ergs (Bloom, Frail, \& Kulkarni 2003).
Pandey et al. (2002) find, in a fit to the optical data through day 21,
that there is a break in the light curve at $t_b=7.6\pm 0.3$ days.
Bloom et al. (2003) interpret this as a jet break; the correction for
collimated flow reduces the $\gamma$-ray energy by a factor of $\sim 40$.
However, Fig. 1 shows that the variability in the light curve and
the late light curve points make a clear designation of the jet
break difficult.
In a wind model, the jet break evolves slowly (Kumar \& Panaitescu 
2000), which makes any determination of a jet break from afterglow
data more uncertain. 
%



The mass loss rate we deduced is typical for a Wolf-Rayet type wind.
There are other indications that GRB 021004 may be interacting with 
a Wolf-Rayet type wind. Wolf-Rayet winds are thought to be clumpy 
(e.g., Hamann \& Koesterke 1998), 
and the clumpiness may provide 
an explanation for the prominent
bumps on the R-band light curve of GRB 021004 (Lazzati et al. 2002; 
Nakar et al. 2002; Heyl \& Perna 2003). In addition, there are multiple 
absorption components in the spectrum of the afterglow, separated by
speeds up to 3000~km~s$^{-1}$. These components could come from 
substructures in a Wolf-Rayet wind (Mirabal et al. 2002; Schaefer
et al. 2002). 
Interestingly, GRB 990510, which is best modeled by interaction with
a constant density medium (CL00; Panaitescu \& Kumar 2002), shows a
smoothly evolving optical afterglow (Stanek et al. 1999).

GRB 021004 differs from the other two GRBs with detected early afterglows 
(GRBs 990123 and 021211) in several ways: it has a slow decay in the 
R-band light curve followed by a steepening rather than a steep decline 
followed by a flattening (which occurs at a much earlier time than the
break in GRB 021004), a higher optical flux at late times after the 
break, and a bright, long-lived radio afterglow. The early emission 
from GRB 990123 was convincingly interpreted as coming from the reverse 
shock of a blast wave expanding into an ISM (Sari \& Piran 1999; 
M\'esz\'aros \& Rees 1999), and the late-time afterglow data are 
consistent with an ISM model (e.g., Panaitescu \& Kumar 2002). The 
afterglow of GRB 021211 resembles that of GRB 990123, and was 
interpreted similarly (Fox et al. 2003a; Li et al. 2003; Wei 2003). 
The rate of initial decline in the R-band flux of approximately $t^{-2}$ 
(GRB 990123) or shallower (GRB 021211) is difficult to reproduce in 
the reverse shock of a wind model. The available data on the three 
early afterglows therefore appear to point to two types of GRB ambient 
environments: ISM and a stellar wind. These two types of environments 
are also inferred in the detailed modeling of a larger number of 
afterglows at later times (CL00; Panaitescu \& Kumar 2002; Price et al. 2002).

The evidence for wind and constant density environments in the early
afterglow evolution supports the hypothesis that these represent different
progenitors (Chevalier \& Li 1999; CL00) because a small radial distance from the star
is explored ($\la 10^{17}$ cm) where a freely expanding
wind is expected.
A prediction of the CL99 scenario is that wind interaction
should be correlated with supernova light; this hypothesis was supported
by GRB 011121 (Price et al. 2002).
However, GRB 021211, which has an early afterglow indicating ISM interaction,
shows some evidence for a supernova-like bump 
in the light curve (Fruchter et al. 2002;
Testa et al. 2003).
GRB 020405 is another case of apparent ISM interaction and a supernova-like
bump (Berger et al. 2003).
Spectra confirming the supernova nature of light curve bumps, as well
as detailed afterglow observations, are needed to clarify the situation.
%
Long term 
monitoring of radio afterglows will be crucial in testing the wind 
model of early afterglows
(by examining the evolution of $\nu_m$), and in determining the wind density. 
Another prediction of the wind model is that the early 
optical emission before the break ($t< t_m$) should have the 
spectrum $F_{\nu}\propto \nu^{-1/2}$, which is a flatter spectrum 
than is typically observed in optical afterglows.
In addition, the break is chromatic, occurring at a later time for
a longer wavelength, which can be tested with densely sampled IR
observations. 


\acknowledgments
Support for this work was provided in part by NASA.

\clearpage

\begin{figure}

\includegraphics[scale=0.60]{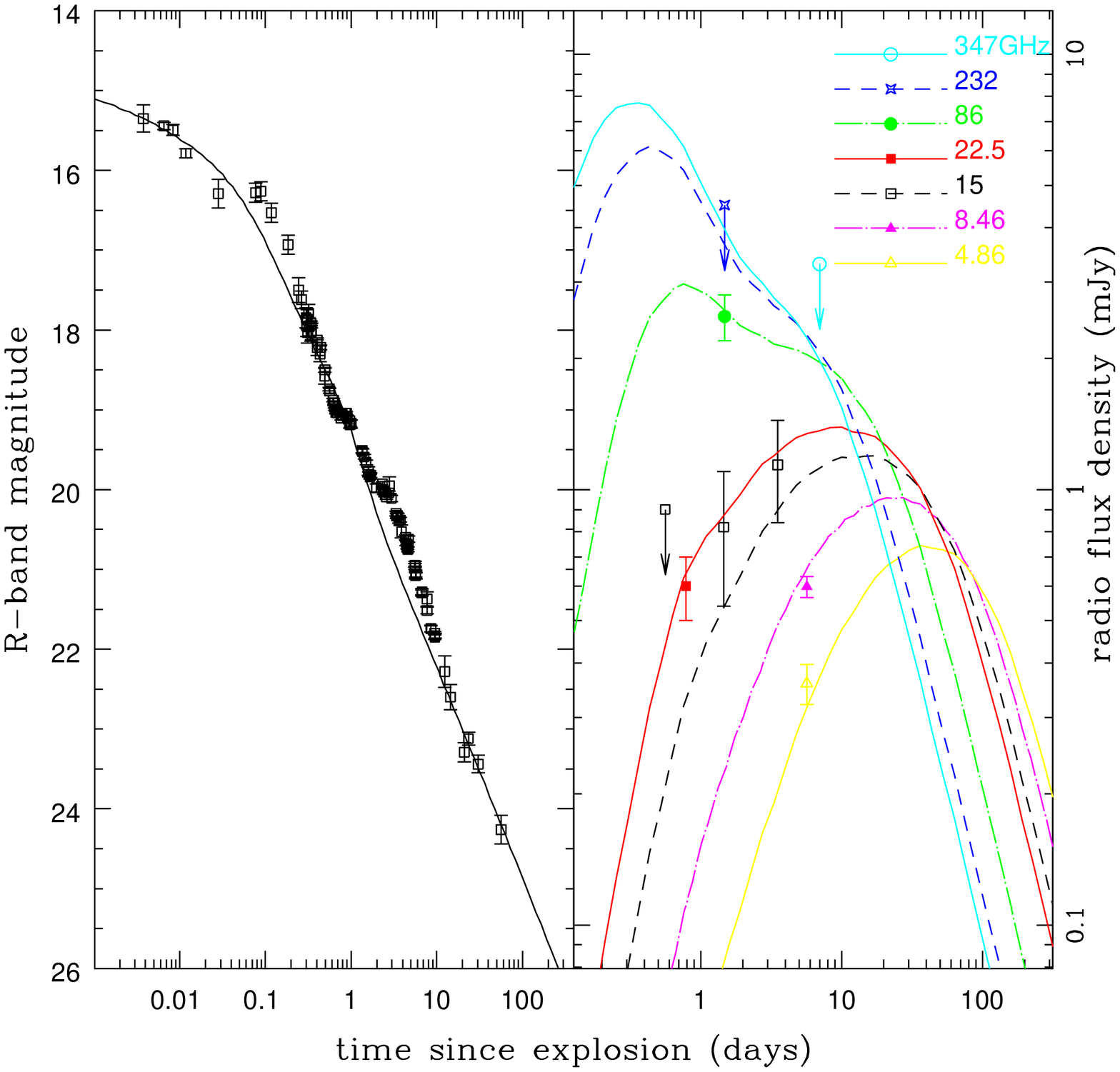}

\caption{Wind interaction model for the afterglow of GRB 021004. 
The optical data are taken from the papers Fox et al. (2003b), 
Bersier et al. (2003), and Holland et al. (2003) and the GCN 
notices Matsumoto et al. (2002), Weidinger et al. (2002), 
Mirabal et al. (2002) and Fatkhullin et al. (2002), with the 
modest amount of Galactic extinction corrected. The radio data 
are taken from Frail \& Berger (2002; 22.5 GHz), Berger, Frail 
\& Kulkarni (2002; 4.86 and 8.46 GHz), Pooley (2002b,c; 15 GHz), 
and Bermer \& Castro-Tirado (2002; 86 GHz). The upper limits at 
15, 232 and 347 GHz are given in Pooley (2002a), Bermer \& 
Castro-Tirado (2002) and Wouterloot et al. (2002). The lines 
are the light curves from the wind model described in the text.  
}

\end{figure}


\clearpage



\begin{thebibliography} {}

\bibitem[]{}
Akerlof, C. W., et al. 1999, Nature, 398, 400


\bibitem[]{}%
Berger, E., Frail, D. A., \& Kulkarni, S. R. 2002, GCN 1613

\bibitem[]{}%
Berger, E., Soderberg, A. M., Frail, D. A., \& Kulkarni, S. R. 2003,
ApJ, submitted (astro-ph/0301634)

\bibitem[]{}
Bersier, D., et al. 2003, ApJ, 584, L43

\bibitem[]{}
Bloom, J. S., Frail, D. A., \& Kulkarni, S. R. 2003, ApJ, submitted (astro-ph/0302210)

\bibitem[]{}%
Bremer, M., \& Castro-Tirado, A. J. 2002, GCN 1590

\bibitem[]{}%
Chevalier, R. A., \& Li, Z.-Y. 1999, ApJ, 520, L29 (CL99)

\bibitem[]{}
Chevalier, R. A., \& Li, Z.-Y. 2000, ApJ, 536, 195 (CL00)

\bibitem[]{}
Chornock, R., \& Filippenko, A. V. 2002, GCN 1605

\bibitem[]{}
Fatkhullin, T. A., Komarova,  V. N., Moiseev,  A. V. 2002, GCN 1717 

\bibitem[]{}%
Fox, D. W., et al. 2003a, ApJ, 586, L5 

\bibitem[]{}%
Fox, D. W., et al. 2003b, Nature, 422, 284


\bibitem[]{}%
Frail, D. A., \& Berger, E. 2002, GCN 1574


\bibitem[]{}
Frail, D. A., Waxman, E., \& Kulkarni, S. R.
2000, \apj, 537, 191

\bibitem[]{}%
Fruchter, A., et al. 2002, GCN 1781


\bibitem[]{}
Granot, J.,  \& Sari, R. 2002, ApJ, 568, 820 

\bibitem[]{}
Hamann, W.-R., \& Koesterke, L. 1998, A\&A, 335, 1003

\bibitem[]{}
Heyl, J. S., \& Perna, R. 2003, ApJ, 586, L13

\bibitem[]{}
Holland, S. T., et al. 2002, AJ, submitted (astro-ph/0211094)


\bibitem[]{}%
Kobayashi, S., \& Zhang, B. 2003, ApJ, 582, L75


\bibitem[]{}
Kumar, P., \& Panaitescu A. 2000, \apj, 541, L9

\bibitem[]{}
Kumar, P., \& Panaitescu, A. 2000, ApJ, 541, L51


\bibitem[]{}
Lazzati, D., Rossi, E., Covino, S., Ghisellini, G., \& Malesani, D. 2002,
A\&A, 396, L5


\bibitem[]{}%
Li, W., Filippenko, A. V., Chornock, R., \& Jha, S. 2003, ApJ, 586, L9


\bibitem[]{}
Li, Z.-Y., \& Chevalier, R. A. 2001, \apj, 551, 940


\bibitem[]{}
Matheson, T., et al. 2003, ApJ, 582, L5

\bibitem[]{}
Matsumoto, K., et al. 2002, GCN 1594 


\bibitem[]{}
M\'esz\'aros, P. \& Rees, M. J. 1997, \apj, 476, 232

\bibitem[]{}
M\'esz\'aros, P. \& Rees, M. J. 1999, MNRAS, 306, L39

\bibitem[]{}%
Mirabal, N., Halpern, J. P., Chornock, R., \& Filippenko, A. V. 2002, GCN 1618

\bibitem[]{}
Nakar, E., Piran, T., \& Granot, J. 2002, New Astr., submitted (astro-ph/0210631)

\bibitem[]{}
Panaitescu, A., \& Kumar, P. 2000, ApJ, 543, 66

\bibitem[]{}
Panaitescu, A., \& Kumar, P. 2002, ApJ, 571, 779



\bibitem[]{}%
Pandey, S. B., et al. 2002, BASI, submitted (astro-ph/0211108)

\bibitem[]{}%
Pooley, G. 2002a, GCN 1575

\bibitem[]{}%
Pooley, G. 2002b, GCN 1588

\bibitem[]{}%
Pooley, G. 2002c, GCN 1604

\bibitem[]{}
Price, P. A., et al. 2002, ApJ, 572, L51

\bibitem[]{}%
Rhoads, J. E. 1997, ApJ, 487, L1



\bibitem[]{}%
Sako, M., \& Harrison, F. A. 2002a, GCN 1624

\bibitem[]{}%
Sako, M., \& Harrison, F. A. 2002b, GCN 1716


\bibitem[]{}
Sari, R. \& Piran, T. 1999, ApJ, 517, L109


\bibitem[]{}%
Sari, R., Piran, T., \& Halpern, J. P. 1999,
\apj, 519, L17

\bibitem[]{}%
Sari, R., Piran, T., \& Narayan, R. 1998,
\apj, 497, L17

\bibitem[]{}%
Schaefer, B. E., et al. 2002, ApJ, submitted (astro-ph/0211189)

\bibitem[]{}%
Shirasake, Y., et al. 2002, GCN 1565


\bibitem[]{}%
Stanek, K. Z., Garnavich, P. M., Kaluzny, et al. 1999,
\apj, 522, L39

\bibitem[]{}%
Testa, V., et al. 2003, GCN 1821





\bibitem[]{}%
Wei, D. M. 2003, preprint (astro-ph/0301345)

\bibitem[]{}%
Weidinger, M., et al. 2002, GCN 1573




\bibitem[]{}%
Wouterloot, J., et al. 2002, GCN 1627


\end{thebibliography}
\end{document}